# Application of classical statistical mechanics to multifractals and dynamical systems

(Short title: Statmech of multifractals)

S.G. Abaimov[1]

[1]University of California-Davis, Davis, 95616, CA, USA

E-mail: sgabaimov@ucdavis.edu

**Abstract.** Classical, self-consistent theory of statistical mechanics was developed for the thermodynamic and conservative Hamiltonian systems. Later there were many attempts (Sinai-Bowen-Ruelle's temperature, Tsallis' non-extensive theory) to apply similar formalism to non-Hamiltonian dynamical systems. Although these theories reveal aspects of complex behavior, they have limited applicability. This paper applies the classical Gibbs-Boltzmann statistical mechanics to complex systems such as i.i.d. processes, multifractals, and non-Hamiltonian dynamical systems with strange attractors. The effective thermolization of stochastic noise in the system is introduced and the formalism of a ruling (governing, free energy) potential is developed.

## 1. Introduction

The formalism of Gibbs-Boltzmann statistical mechanics was developed for conservative Hamiltonian systems. Later attempts have been made to apply this formalism to non-Hamiltonian dynamical systems. Particularly, Sinai [1], Bowen [2], and Ruelle [3] have introduced an artificial temperature parameter. Let's follow their approach and assume that we have some stationary distribution of probabilities $p_\alpha$ over a set of microstates. Everywhere further in this paper index $\alpha$ will enumerate microstates in system phase space. Probabilities $p_\alpha$ are normalized: $\sum_\alpha p_\alpha = 1$. To each microstate $\alpha$ with the probability $p_\alpha$ we can assign effective energy $E_\alpha = -\ln p_\alpha$. Then $p_\alpha = e^{-E_\alpha}$. Partition function of the system $z \equiv \sum_\alpha e^{-E_\alpha}$ is a normalization constant of probabilities $z = \sum_\alpha p_\alpha = 1$. It always equals unity as the temperature is naturally absent in the system.

We can introduce an artificial temperature $\frac{1}{T} \equiv \beta$ by including it as a parameter into the definition of the partition function: $z_\beta = \sum_\alpha e^{-\beta E_\alpha}$. Then we can introduce a renormalized probability distribution as $w_\alpha \equiv \frac{1}{z_\beta} p_\alpha^\beta = \frac{1}{z_\beta} e^{-\beta E_\alpha}$. In statistical mechanics the averaged physical properties of a canonical ensemble can be found as derivatives of partition function. Particularly, for the averaged energy we have $\frac{\partial \ln z_\beta}{\partial \beta} = -\sum_\alpha E_\alpha \frac{1}{z_\beta} e^{-\beta E_\alpha} = -\sum_\alpha E_\alpha w_\alpha = -<E>_w$ where $<...>_w$ is the ensemble average with probabilities $w_\alpha \equiv \frac{1}{z_\beta} p_\alpha^\beta$. The system could be investigated for different values of $\beta$ and a complex behavior could be found. This looks like a direct analogy with the classical statistical mechanics although *in situ* it is not. Indeed, in physics a scientist usually operates with averaged measurable quantities. Averaging with the probability distribution $w_\alpha$ means averaging with probabilities $p_\alpha^\beta$ which is not the measurable averaging of a physical quantity. Only in the limit $\beta \to 1$ we have physically relevant results. However, this is the trivial case of the absence of the artificially introduced temperature parameter. Another similar way to develop the non-extensive statistical mechanics was suggested by Tsallis [see, e.g., 4]. But again the main limitation is the inability to obtain the measurable average of a physical quantity. Both Sinai-Bowen-Ruelle's theory and non-extensive theory are based on the introduction of an artificial q-exponent for the measure. The most fruitful application of this technique has been developed in the theory of multifractals. *q*-partition function, Lipschitz-Holder exponent *α*, or mass exponents *τ*(*q*) are all *in situ* parameterized by the artificial temperature.

This paper follows a classical approach for complex non-thermal phenomena. Particularly, the stochastic noise prescribed to a system as an external boundary constraint is described in terms of the effective temperature parameters thermolizing the system. Then the ruling (governing, free energy) potential is introduced and its behavior is investigated. To illustrate this theory, three different systems are considered: an i.i.d. process, a multifractal, and a dynamic system with a strange attractor.

## 2. Models

Three models are considered: an i.i.d. process, a multifractal, and a dynamical system with a strange attractor. As a model of an independent identically distributed (i.i.d.) process a biased coin with *n* possible outcomes is chosen. So, the alphabet [$A_1,…, A_n$] has size *n*. Let $p_1,…,p_n$ be the probabilities of symbols $A_1,…, A_n$ respectively in the stationary distribution with the constraint $\sum_{i=1}^{n} p_i = 1$. As an $N^{th}$ iteration of the process we consider the chain of *N* symbols. A particular realization of this chain has $k_1,…, k_n$ symbols $A_1,…, A_n$ respectively with the constraint $\sum_{i=1}^{n} k_i = N$. By the vector $\vec{k} = \|k_1 \quad … \quad k_n\|$ we will denote the symbol content of a particular realization. As a microstate α we will understand a particular realization of N symbols. Probability of this microstate to occur is $p_1^{k_1} \cdot ... \cdot p_n^{k_n}$. As a macrostate $\vec{k}$ we will usually understand a union of all possible microstates corresponding to the given vector $\vec{k}$. If another definition of the macrostate shall be used it will be clearly specified.

As a model for a multifractal we consider a 1-dimentional Cantor set. Each branch of iteration N, measure P, and length Q generates n branches of iteration N + 1 with measures $P p_1, …, P p_n$ and lengths $Q q_1, …, Q q_n$. Again there is constraint $\sum_{i=1}^{n} p_i = 1$ for the measure redistribution. A particular branch of the iteration N has been formed through the way $k_1, …, k_n$ with the constraint $\sum_{i=1}^{n} k_i = N$ where $k_i$ is the number of times when the measure has been redistributed with the probability $p_i$. A microstate α is a particular branch of iteration N with measure $p_1^{k_1} \cdot ... \cdot p_n^{k_n}$ and length $q_1^{k_1} \cdot ... \cdot q_n^{k_n}$. The definition of a macrostate is similar to the previous case and if all $p_i$ and $q_i$ are different is *in situ* the definition of a multifractal.

As a model of a dynamical system we will consider a map with a strange attractor. For this attractor the iteration N of the cylinder partition (N-cylinder) is assumed to be constructed. Mixing and ergodic properties of the attractor are assumed to provide a stationary measure distribution of the symbolic process over this partition with the probability $p_1^{k_1} \cdot ... \cdot p_n^{k_n}$ in a partition cell. Here vector $\vec{k}$ again denotes the symbolic way (content) by which this cell has been formed. A microstate α is a particular partition cell of the iteration N with the measure $p_1^{k_1} \cdot ... \cdot p_n^{k_n}$. The definition of a macrostate is similar to the previous cases.

As an objection for the map here could be the fact that *in situ* the process over the partition has memory and is not an i.i.d. process. Therefore to look only at its averaged stationary distribution could be an oversimplification. However, Hamiltonian thermal systems like an ideal gas model are all *in situ* Markov processes too. In spite of this fact the Gibbs-Boltzmann statistical mechanics which is in fact the i.i.d. mechanics is perfectly applicable for these systems. The reason is the quick ergodic mixing which looses any memory of the quick processes in the stationary distribution of the system.

All three systems are not thermodynamic and thermal fluctuations are absent in them. Instead, the model stochasticity of the systems $p_1, \ldots, p_n$ is prescribed as an external constraint. Therefore these systems could be named as the systems with the prescribed 'stochastical' noise. Our purpose further is to describe this noise in terms of the effective temperature parameters which will thermolize the systems.

We should remember that there are additional constraints $\sum_{i=1}^{n} p_i = 1$ and $\sum_{i=1}^{n} k_i = N$. Further as the dependent variables we consider $p_n$ and $k_n$ and will exclude them from equations. For any of three models above we have

$$N_{\vec{k}} = \frac{N!}{k_1! \ldots k_n!} = \frac{N!}{k_1! \ldots k_{n-1}!(N - k_1 - \ldots - k_{n-1})!} \tag{1}$$

as a number of microstates corresponding to the macrostate $\vec{k}$. The probability $w_{\vec{k}}^{equil}$ of each of these microstates is

$$w_{\vec{k}}^{equil} = p_1^{k_1} \ldots p_n^{k_n} = p_1^{k_1} \ldots p_{n-1}^{k_{n-1}} (1 - p_1 - \ldots - p_{n-1})^{N - k_1 - \ldots - k_{n-1}} \tag{2}$$

This probability is dictated by the model parameters $p_1, \ldots, p_n$ prescribed as an external stochasticity constraint. This constraint is similar to the temperature of the external medium in canonical ensemble. It dictates the equilibrium distribution of probabilities but the system actually can realize itself in a non-equilibrium state with any other probability distribution. Only equilibrium state is dictated by the external constraint therefore we used abbreviation 'equil' to emphasize that this probability distribution corresponds to the equilibrium with the external constraint. *E.g.*, we could think of a biased coin. This process can realize any, even highly improbable sequences of symbols from its alphabet and the given probability distribution $p_i$ as the external model constraint describes only the most probable chains of symbols.

Each of $N_{\vec{k}}$ microstates of macrostate $\vec{k}$ has probability $w_{\vec{k}}^{equil}$. The total probability of the macrostate $\vec{k}$ is $W^{equil}(\vec{k}) = N_{\vec{k}} w_{\vec{k}}^{equil}$. This is the probability for the macrostate $\vec{k}$ to be observed in equilibrium with the external stochasticity constraint.

In the thermodynamic limit $N \to +\infty$ we can approximate eq. (1) by

$$N_{\vec{k}} \approx_{\ln} \frac{1}{\left(\frac{k_1}{N}\right)^{k_1} \ldots \left(\frac{k_{n-1}}{N}\right)^{k_{n-1}} \left(\frac{N - k_1 - \ldots - k_{n-1}}{N}\right)^{N - k_1 - \ldots - k_{n-1}}} \quad (3)$$

where the symbol "$\approx_{\ln}$" means that all power-law multipliers are neglected in comparison with the exponential dependence on $N$. Everywhere further the symbol "$\approx_{\ln}$" will mean the accuracy of exponential dependence neglecting all power-law dependences. For the logarithm of such equations we will use the symbol "$\approx$".

For the $W^{equil}(\vec{k})$ we have

$$W^{equil}(\vec{k}) \approx_{\ln} \left(\frac{p_1}{k_1/N}\right)^{k_1} \cdots \left(\frac{p_{n-1}}{k_{n-1}/N}\right)^{k_{n-1}} \left(\frac{1-p_1-\ldots-p_{n-1}}{(N-k_1-\ldots-k_{n-1})/N}\right)^{N-k_1-\ldots-k_{n-1}} \tag{4}$$

## 3. Equation of state

$W^{equil}(\vec{k})$ is the probability distribution for the system in equilibrium with the external stochasticity constraint. To obtain the averaged values of measurable quantities in this state (here often the term 'equilibrium' is used too but to avoid confusion we will use superscript $^0$) it is necessary to maximize $W^{equil}(\vec{k})$ over all possible values of $\vec{k}$. Both functions $N_{\vec{k}}$ and $w_{\vec{k}}^{equil}$ depend exponentially on $N$ which is infinite in the thermodynamic limit. Therefore these functions have very rapid change with the change of $\vec{k}$ and the maximum of $W^{equil}(\vec{k})$ is very sharp. To find the maximum it is necessary to find where derivatives $\frac{\partial W^{equil}}{\partial k_{i=1,\ldots,n-1}}(\vec{k})$ or $\frac{\partial \ln W^{equil}}{\partial k_{i=1,\ldots,n-1}}(\vec{k})$ equal to zero. After simple algebra we obtain that the equilibrium value of $\vec{k}$ is

$$\vec{k}^0 = N \| p_1 \quad \ldots \quad p_n \| \tag{5}$$

as it could be expected. Eq. (5) is the equation of the equilibrium state.

It is easy to find that the second derivatives of the function $\ln W^{equil}(\vec{k})$ at the point of maximum (5) equal $\left.\frac{\partial^2 W^{equil}}{\partial k_{i=1,\ldots,n-1} \partial k_{j=1,\ldots,n-1}}\right|_{\vec{k}^0} = -\frac{1}{N}\frac{\delta_{ij}}{\sqrt{p_i p_j}} - \frac{1}{N}\frac{1}{p_n}$. The second derivatives are negative therefore the obtained extremum is indeed a maximum. Also because the maximum is very narrow we can approximate its curvature by the bilinear descent and find that the width of the maximum is of the order of $|\delta\vec{k}| \propto \sqrt{N}$. Fluctuations of $\vec{k}$ in equilibrium have an order of the maximum width and therefore relative fluctuations are inversely proportional to the square root of $N$: $\frac{|\delta \vec{k}|}{|<\vec{k}>|} \propto \frac{1}{\sqrt{N}}$. Indeed, the logarithm of the probability $W^{equil}(\vec{k})$ of fluctuations in the vicinity of the maximum can be approximated by the bilinear dependence

$$\ln W^{equil}(\vec{k}) = \ln W^{equil}(\vec{k}^0) + \frac{1}{2}\sum_{i,j=1}^{n-1}\frac{\partial^2 \ln W^{equil}}{\partial k_i \partial k_j}(\vec{k}^0) \cdot \Delta k_i \Delta k_j$$ or

$$W^{equil}(\vec{k}) \propto \exp\left(-\frac{1}{2N}\sum_{i,j=1}^{n-1}\left(\frac{\delta_{ij}}{\sqrt{p_i p_j}} + \frac{1}{p_n}\right) \cdot \Delta k_i \Delta k_j\right)$$ and fluctuations are distributed in accordance with the Gaussian distribution. So, the relative fluctuations of $\vec{k}$ are inversely proportional to the square root of $N$ ($N$ is infinite in the thermodynamic limit). Therefore the maximum is indeed very narrow. Ideally, to obtain any quantity in equilibrium, we must average it over all microstates: $f^{equil} = \sum_\alpha w_\alpha^{equil} f_\alpha = \sum_{\vec{k}} N_{\vec{k}} w_{\vec{k}}^{equil} f_{\vec{k}}$. The fact that the maximum is very narrow gives us a possibility to calculate all quantities averaged over the whole range of $\vec{k}$ as their values at the point of the maximum: $f^{equil} \approx f(\vec{k}^0)$. For example, the averaged parameter $\vec{k}$ in the equilibrium equals to its value in the maximum: $\vec{k}^{equil} \approx \vec{k}^0$.

For the equilibrium probability distribution the entropy is $S^{equil} \equiv -\sum_\alpha w_\alpha^{equil} \ln w_\alpha^{equil} = -\sum_{\vec{k}:k_1+\ldots+k_n=N} N_{\vec{k}} w_{\vec{k}}^{equil} \ln w_{\vec{k}}^{equil}$. Substituting eq. (2) we get

$$S^{equil} = -\sum_{\vec{k}:k_1+\ldots+k_n=N} N_{\vec{k}} w_{\vec{k}}^{equil}\{k_1 \ln p_1 + \ldots + k_n \ln p_n\} = -k_1^{equil} \ln p_1 - \ldots - k_n^{equil} \ln p_n = -\ln w_{\vec{k}^{equil}} \approx -\ln w_{\vec{k}^0}.$$ As

$N_{\vec{k}^0} \approx_{\ln} 1/w_{\vec{k}^0}$ therefore $S^{equil} \approx \ln N_{\vec{k}^0}$. The number of microstates $\Delta \Gamma$ in the range of the width of the maximum $\delta\vec{k}$ equals to the product of the number of microstates $N_{\vec{k}}$ for the given value of $\vec{k}$ and the number of different $\vec{k}$ in the region $\delta\vec{k}$ of the maximum: $\Delta\Gamma \propto N_{\vec{k}^0} \frac{|\delta\vec{k}|}{|\Delta\vec{k}|}$. Here $|\delta\vec{k}| \propto \sqrt{N}$ is the width of the maximum and $|\Delta\vec{k}| \propto 1$ is the unit step of increment of $\vec{k}$. Again, neglecting the power-law dependences on $N$ in comparison with the exponential dependence we obtain $\Delta\Gamma \approx_{\ln} N_{\vec{k}^0}$. Therefore $S^{equil} \approx \ln N_{\vec{k}^0} \approx \ln \Delta\Gamma$. We have obtained a fundamental result: the definition of entropy as the negative logarithm of microstate probability averaged over all microstates $S \equiv -<\ln w_\alpha>$ is equivalent to another definition of the entropy as the logarithm of the number of microstates over which the system presumably realizes itself. Using eq. (4) for the entropy of equilibrium state (5) we obtain $S^{equil} \approx \ln N_{\vec{k}_0} \approx -N \sum_{i=1}^{n} p_i \ln p_i$.

## 4. Temperature and ruling potential

To construct a ruling potential it is necessary to discuss first the derivation of the ruling potential in statistical mechanics. First we consider a thermodynamic system isolated with the given energy $E$. In the energy spectrum of the system some $g(E)$ degenerated levels correspond to this value of energy $E$. Therefore only these $g(E)$ microstates are possible for the isolated system. The equilibrium probability of each microstate is $w_\alpha^{equil} = 1/g(E)$. The entropy of the system in equilibrium is

$$S^{equil} \equiv -\sum_{\alpha=1}^{g(E)} w_\alpha^{equil} \ln w_\alpha^{equil} = -g(E) \frac{1}{g(E)} \ln \frac{1}{g(E)} = \ln g(E) \qquad (6)$$

Here $g(E)$ is the number of microstates $\Delta\Gamma$ over which the system can realize itself with non-zero probability and $S^{equil} = \ln\Delta\Gamma$.

For the isolated system we have to use another definition of a macrostate: we define a non-equilibrium macrostate as a system that can realize itself only at a subset $\Delta g$ of all possible microstates $g(E)$: $\Delta g \subset g(E)$. Then the probability of each microstate for this macrostate is $w_\alpha = 1/\Delta g$. There is already no superscript 'equil' in the probability here because this probability is not in equilibrium with the external constraint $E = \text{const}$. The entropy of the macrostate is $S^{macro}_{\Delta g} \equiv -\sum_{\alpha=1}^{\Delta g} w_\alpha \ln w_\alpha = -\Delta g \frac{1}{\Delta g} \ln \frac{1}{\Delta g} = \ln \Delta g$. The probability of this macrostate in the isolated system (the probability to occur in equilibrium with the constraint $E = \text{const}$) is $W^{equil}_{\Delta g} = \Delta g \cdot w^{equil}_\alpha = \Delta g / g(E)$. This probability is *in situ* the ruling potential that should be maximized. The maximum of $W^{equil}_{\Delta g}$ corresponds to the equilibrium macrostate that occupies all possible microstates: $\Delta g^0 = g(E)$. The entropy of this macrostate equals the entropy of the system at the equilibrium: $S^{macro}_{\Delta g^0 = g(E)} = \ln g(E) = S^{equil}$.

Instead of the actual ruling potential $W^{equil}_{\Delta g}$ that should be maximized we can construct a potential that should be minimized. One of the possible choices is $\Phi_{\Delta g} = -\ln W^{equil}_{\Delta g}$ because minus logarithm is a monotonically decreasing function. So defined potential identically equals to zero at the equilibrium macrostate $\Delta g^0 = g(E)$ as $W^{equil}_{\Delta g^0 = g(E)} = 1$. Another possible choice is $\Phi_{\Delta g} = -\ln(g(E) W^{equil}_{\Delta g})$ where $g(E)$ is a constant which does not influence the behavior of the potential. Now $\Phi_{\Delta g} = -\ln \Delta g = -S^{macro}_{\Delta g}$.

Therefore the negative entropy of the non-equilibrium macrostates plays for the isolated system the role of the ruling potential that should be minimized.

For our systems, instead of a system isolated with energy $E$ we can imagine a system isolated with the particular $\vec{k}$ (a system isolated on a particular multifractal). This system can realize only $N_{\vec{k}}$ microstates given by eq. (4). Probability of each of these microstates is $w_{\vec{k}}^{equil} = 1/N_{\vec{k}}$ (equilibrium with the isolation constraint) and the entropy of the system in the equilibrium is $S^{equil} \equiv -\sum_{\alpha=1}^{N_{\vec{k}}} w_{\alpha}^{equil} \ln w_{\alpha}^{equil} = -N_{\vec{k}} \frac{1}{N_{\vec{k}}} \ln \frac{1}{N_{\vec{k}}} = \ln N_{\vec{k}}$.

For the non-equilibrium macrostate $\Delta N$ we have to use here an alternative definition, different from used in section 2. The non-equilibrium macrostate $\Delta N$ is defined as a macrostate when the system with probabilities $w_{\alpha} = 1/\Delta N$ can realize itself only on $\Delta N$ of all possible $N_{\vec{k}}$ microstates. The entropy of this macrostate is $S_{\Delta N}^{macro} \equiv -\sum_{\alpha=1}^{\Delta N} w_{\alpha} \ln w_{\alpha} = -\Delta N \frac{1}{\Delta N} \ln \frac{1}{\Delta N} = \ln \Delta N$ and the probability of this macrostate $\Delta N$ in the isolated system is $W_{\Delta N}^{equil} = \Delta N / N_{\vec{k}}$. This very probability $W_{\Delta N}^{equil}$ is the ruling potential that has to be maximized. Instead, we can construct the ruling potential that has to be minimized as $\Phi_{\Delta N} = -\ln W_{\Delta N}^{equil}$ or $\Phi_{\Delta N} = -\ln(N_{\vec{k}} W_{\Delta N}^{equil}) = -\ln \Delta N = -S_{\Delta N}^{macro}$. Again the negative entropy of the non-equilibrium macrostates can be chosen as a ruling potential.

Phenomenological approach for an isolated system claims that the entropy of the isolated system can only increase: $\frac{dS}{dt} \geq 0$. We see that it corresponds to the fact that on its way to the equilibrium the system prefers macrostates with higher probability $W_{\Delta N}^{equil}$ (with higher $\Delta N$):

$$\frac{dS_{\Delta N}^{macro}}{dt} \geq 0 \qquad (7)$$

Now we consider the case of a canonical ensemble in statistical mechanics ($N$ = const, $V$ = const, $T$ = const). In the canonical ensemble the temperature of the external medium as an external constraint dictates to the system the equilibrium energy $E^0$ and the equilibrium probability of microstates

$$w_\alpha^{equil} = \frac{1}{z} e^{-\frac{E_\alpha}{T}} \qquad (8)$$

where $z$ is the partition function of the system $z = \sum_\alpha e^{-\frac{E_\alpha}{T}}$. The entropy of the system in equilibrium is $S^{equil} \equiv -\sum_\alpha w_\alpha^{equil} \ln w_\alpha^{equil} = -\sum_E g(E) w_E^{equil} \ln w_E^{equil}$ where the sum over microstates has been substituted by the sum over the values of energy, and $g(E)$ denotes again the degeneration of the energy level $E$. Substituting eq. (8) we obtain $S^{equil} = -\sum_E g(E) w_E^{equil} \left\{ -\ln z - \frac{E}{T} \right\} = \ln z + \frac{E^{equil}}{T}$. Defining Helmholtz free energy $A$ as $A = E - TS$ (both for equilibrium and non-equilibrium states) we obtain $A^{equil} = E^{equil} - TS^{equil} = -T \ln z$. So, the Helmholtz free energy equals $-T \ln z$ only for equilibrium states. The equilibrium probability (8) is $w_\alpha^{equil} = e^{\frac{A^{equil} - E_\alpha}{T}}$. Although the Helmholtz free energy is traditionally defined as $A = E - TS$ we could notice that $T$ is constant and therefore the equivalent definition could be $A = E/T - S$. As we will see below only this alternative definition is relevant for the multi-temperature systems.

We can define a non-equilibrium macrostate as a subset of all microstates corresponding to the given energy $E$ (*i.e.*, as a system isolated with $E$). The number of these microstates is $g(E)$ and their probabilities are $w_E = 1/g(E)$ (for the system is constrained by this macrostate, *i.e.* isolated with the given $E$). The entropy of this macrostate is

$$S_E^{macro} = -g(E)\frac{1}{g(E)}\ln\frac{1}{g(E)} = \ln g(E) \tag{9}$$

and the probability of this macrostate in the canonical ensemble (to occur in equilibrium with the constraint $T$ = const) is $W^{equil}(E) = g(E)w_E^{equil} = g(E)e^{\frac{A^{equil}-E}{T}}$. This very probability function $W^{equil}(E)$ is the ruling potential that should be maximized. Also we can define the ruling potential that has to be minimized as $\Phi(E) = -\ln W^{equil}(E)$. The maximum of $W^{equil}(E)$ is very narrow, the number of energy levels $\Delta\Gamma$ in its width has an order of the degeneration of one of them $g(E^0)$ (again neglecting power-law dependences on $N$ in comparison with the exponential dependence of $g(E)$). But the area under the function $W^{equil}(E)$ has to accumulate its unity value under the maximum. Therefore we can conclude that at the maximum $g(E^0) \approx_{\ln} 1/w_{E^0}^{equil}$ where $E^0$ is the equilibrium value of energy at the maximum. Therefore at the equilibrium macrostate

$$\Phi(E^0) = -\ln W^{equil}(E^0) = -\ln g(E^0)w_{E^0}^{equil} \approx 0 \tag{10}$$

For any equilibrium state this potential identically equals zero. Therefore its derivatives over the equilibrium changes also equal zero identically. Later we will introduce a Helmholtz free energy as a potential whose derivatives could be non-zero. Therefore the criterion to distinguish first and continuous phase transitions strongly depends on the choice of the ruling potential.

The maximum is very narrow and $\ln w_E^{equil}$ is a slowly changing function with a power-law dependence on $N$ in comparison with $g(E)$ and $w_E^{equil}$ with an exponential dependence on $N$. Therefore for the entropy of the system in equilibrium we have $S^{equil} = -\sum_E g(E) w_E^{equil} \ln w_E^{equil} \approx -\ln w_{E^0}^{equil} \sum_E g(E) w_E^{equil} = -\ln w_{E^0}^{equil} \approx \ln g(E^0) \approx \ln \Delta\Gamma$. The entropy of the equilibrium macrostate is $S_{E^0}^{macro} = \ln g(E^0)$ and therefore again the entropy of the system at the equilibrium equals to the entropy of the most probable macrostate and equals the logarithm of the number of microstates over which the system can realize itself.

Also we can define the ruling potential that has to be minimized as $\Phi(E) = -\lambda_1 \ln(W^{equil}(E)\lambda_2)$ where $\lambda_1$ and $\lambda_2$ are some positive constants. Choosing these constants to be $\lambda_1 = T$ and $\lambda_2 = z$ we obtain

$$\Phi(E) = -T \ln W^{equil}(E) + A^{equil} = -T \ln g(E) - T \ln e^{\frac{A^{equil}-E}{T}} + A^{equil} = -TS_E^{macro} + E = A \qquad (11)$$

Ruling potential (11) corresponds to the Helmholtz free energy for equilibrium and non-equilibrium states. While potential (10) is identically equal to zero for any equilibrium state and therefore its equilibrium derivatives are zero too, the Helmholtz free energy (11) could have complex behavior of its derivatives characterizing the order of the possible phase transition in the system. Therefore the classification of orders of phase transitions significantly depends on the choice of the ruling potential.

At the maximum of $W^{equil}(E)$ we have $\frac{dW^{equil}}{dE}(E^0) = 0$ or $\frac{d\ln W^{equil}}{dE}(E^0) = 0$. For

$\ln W^{equil}(E) = \ln g(E) + \frac{A^{equil} - E}{T}$ we obtain

$$\frac{1}{T} = \frac{d\ln g(E)}{dE}\bigg|_{E^0} = \frac{d\ln g(E^0)}{dE^0} \qquad (12)$$

at the equilibrium state. Often this equation is used as a definition of temperature. As both the entropy of the macrostate $S_E^{macro} = \ln g(E)$ and the equilibrium entropy $S^{equil} \approx \ln g(E^0)$ have the same functional dependence on $E$ and $E^0$ respectively we obtain $\frac{1}{T} = \frac{dS_E^{macro}}{dE}\bigg|_{E^0} \approx \frac{dS^{equil}}{dE^0}$. This is the energy-balance equation $dE^0 = TdS^{equil}$. For non-equilibrium states instead, the increment of energy equals to the amount of heat received by the system $dE = \delta Q^{\leftarrow}$ where in general $\delta Q^{\leftarrow} < TdS$.

Imagine now a system in canonical ensemble during its evolution over non-equilibrium macrostates $E$ on its way to the equilibrium. Each macrostate $E$ could be thought as a system isolated with the energy $E$. Therefore the increase of the entropy in the system in accordance with eq. (7) must be higher than the increase of the entropy produced only by the change of macrostates (9): $dS_{\Delta N(E)}^{macro} \geq dS_E^{macro} = d\ln g(E)$. In the vicinity of the maximum of $W^{equil}(E)$ we have eq. (12) and $dS \geq d\ln g(E) = dE/T$ or $dA \equiv d(E - TS) = dE - TdS \leq 0$. Therefore we have confirmed that the Helmholtz free energy is a ruling potential in the case of the canonical ensemble.

The behavior of our systems is analogous to the behavior of the canonical ensemble, and parameter $\vec{k}$ plays a role of the energy $E$. Indeed, the equilibrium probability of microstates assigned *a priori* by eq. (2) equals

$$w_{\vec{k}}^{equil} = \exp(k_1 \ln p_1 + \ldots + k_{n-1} \ln p_{n-1} + (N - k_1 - \ldots - k_{n-1}) \ln p_n) = \frac{1}{z} e^{-\frac{k_1}{T_1} - \ldots - \frac{k_{n-1}}{T_{n-1}}} \qquad (13)$$

where $z = p_n^{-N}$ and $\beta_i \equiv \frac{1}{T_i} \equiv \ln \frac{p_n}{p_i}$, $i = 1, \ldots, n-1$. It is easy to verify that $z$ is again the partition function of the system $z = \sum_{\vec{k}: k_1 + \ldots + k_n = N} N_{\vec{k}} e^{-\frac{k_1}{T_1} - \ldots - \frac{k_{n-1}}{T_{n-1}}}$. Parameters $T_i$ play the roles of temperatures prescribed by external constraints $p_i$. Here, instead of defining the external stochasticity constraints as the prescribed $p_i$, we alternatively can define these constraints as the prescribed constant temperature constraints $T_i$ = const for the system.

The entropy of the system in equilibrium is $S^{equil} \equiv -\sum_\alpha w_\alpha^{equil} \ln w_\alpha^{equil} = -\sum_{\vec{k}: k_1 + \ldots + k_n = N} N_{\vec{k}} w_{\vec{k}}^{equil} \ln w_{\vec{k}}^{equil}$ where the sum over microstates has been substituted by the sum over the values of $\vec{k}$ and $N_{\vec{k}}$ is given by eq. (4). Substituting eq. (13) we obtain $S^{equil} = -\sum_{\vec{k}: k_1 + \ldots + k_n = N} N_{\vec{k}} w_{\vec{k}}^{equil} \left\{ -\ln z - \frac{k_1}{T_1} - \ldots - \frac{k_{n-1}}{T_{n-1}} \right\} = \ln z + \frac{k_1^{equil}}{T_1} + \ldots + \frac{k_{n-1}^{equil}}{T_{n-1}}$.

Defining the Helmholtz free energy $A$ as $A \equiv \frac{k_1}{T_1} + \ldots + \frac{k_{n-1}}{T_{n-1}} - S$ we obtain $A^{equil} = \frac{k_1^{equil}}{T_1} + \ldots + \frac{k_{n-1}^{equil}}{T_{n-1}} - S^{equil} = -\ln z$. Therefore for the equilibrium states the Helmholtz free energy is the negative logarithm of the partition function. Then the equilibrium probability (13) is $w_{\vec{k}}^{equil} = e^{A^{equil} - \frac{k_1}{T_1} - \ldots - \frac{k_{n-1}}{T_{n-1}}}$.

We can define a non-equilibrium macrostate as a subset of all microstates corresponding to the given $\vec{k}$ (*i.e.*, as a system isolated with $\vec{k}$). The number of these microstates is $N_{\vec{k}}$ and their probabilities are $w_{\vec{k}} = 1/N_{\vec{k}}$. The entropy of this macrostate is

$$S_{\vec{k}}^{macro} = -N_{\vec{k}} \frac{1}{N_{\vec{k}}} \ln \frac{1}{N_{\vec{k}}} = \ln N_{\vec{k}} \qquad (14)$$

and the probability for this macrostate in equilibrium with the external stochasticity constraint is $W^{equil}(\vec{k}) = N_{\vec{k}} w_{\vec{k}}^{equil} = N_{\vec{k}} e^{A^{equil} - \frac{k_1}{T_1} - \ldots - \frac{k_{n-1}}{T_{n-1}}}$. This very probability function $W^{equil}(\vec{k})$ is the ruling potential that should be maximized. Also we can define the ruling potential that has to be minimized as $\Phi(\vec{k}) = -\ln W^{equil}(\vec{k})$. The maximum of $W^{equil}(\vec{k})$ is very narrow, the number $\Delta\Gamma$ of microstates in its range has an order of the number of microstates $N_{\vec{k}^0}$ corresponding to one particular $\vec{k}$ in this range (again neglecting power-law dependences on $N$ in comparison with the exponential dependence of $N_{\vec{k}}$). But the area under the function $W^{equil}(\vec{k})$ has to accumulate its unity value under the maximum. Therefore we can conclude that at the maximum $N_{\vec{k}^0} \approx_{\ln} 1/w_{\vec{k}^0}^{equil}$ where $\vec{k}^0$ is the equilibrium value of $\vec{k}$ given by eq. (5). Also this result could be verified directly. Therefore at the equilibrium macrostate $\Phi(\vec{k}^0) = -\ln W^{equil}(\vec{k}^0) = -\ln N_{\vec{k}^0} w_{\vec{k}^0}^{equil} \approx 0$.

As the maximum is very narrow for the entropy of the system in equilibrium we have $S^{equil} = -\sum_{\vec{k}:k_1+\ldots+k_n=N} N_{\vec{k}} w_{\vec{k}}^{equil} \ln w_{\vec{k}}^{equil} \approx -\ln w_{\vec{k}^0}^{equil} \sum_{\vec{k}:k_1+\ldots+k_n=N} N_{\vec{k}} w_{\vec{k}}^{equil} = -\ln w_{\vec{k}^0}^{equil} \approx \ln N_{\vec{k}^0} \approx \ln \Delta\Gamma$. The entropy of the equilibrium macrostate is $S_{\vec{k}^0}^{macro} = \ln N_{\vec{k}^0}$. Therefore again the entropy of the system in the equilibrium equals the entropy of the most probable macrostate and equals the logarithm of the number of microstates over which the system can realize itself.

Also we can define the ruling potential that has to be minimized as $\Phi(\vec{k}) = -\lambda_1 \ln(W^{equil}(\vec{k})\lambda_2)$ where $\lambda_1$ and $\lambda_2$ are some positive constants. Choosing these constants to be $\lambda_1 = 1$ and $\lambda_2 = z$ we obtain

$$\Phi(\vec{k}) = -\ln W^{equil}(\vec{k}) + A^{equil} = -\ln N_{\vec{k}} - \ln e^{A^{equil} - \frac{k_1}{T_1} - \ldots - \frac{k_{n-1}}{T_{n-1}}} + A^{equil} = -S_{\vec{k}}^{macro} + \frac{k_1}{T_1} + \ldots + \frac{k_{n-1}}{T_{n-1}} \equiv A.$$

Therefore now the ruling potential corresponds to the Helmholtz free energy for equilibrium and non-equilibrium states.

At the maximum of $W^{equil}(\vec{k})$ we have $\frac{\partial W^{equil}}{\partial k_{i=1,\ldots,n-1}}(\vec{k}^0) = 0$ or $\frac{\partial \ln W^{equil}}{\partial k_{i=1,\ldots,n-1}}(\vec{k}^0) = 0$. For

$\ln W^{equil}(\vec{k}) = \ln N_{\vec{k}} + A^{equil} - \frac{k_1}{T_1} - \ldots - \frac{k_{n-1}}{T_{n-1}}$ we can write that

$$\frac{1}{T_i} \equiv \beta_i = \left.\frac{\partial \ln N_{\vec{k}}}{\partial k_{i=1,\ldots,n-1}}\right|_{\vec{k}^0} = \frac{\partial \ln N_{\vec{k}^0}}{\partial k^0_{i=1,\ldots,n-1}} \qquad (15)$$

at the equilibrium state. This equation could be used as a definition of the temperature parameters. As both the entropy of macrostate $S_{\vec{k}}^{macro} = \ln N_{\vec{k}}$ and the equilibrium entropy $S^{equil} \approx \ln N_{\vec{k}^0}$ have the same functional dependence on $\vec{k}$ and $\vec{k}^0$ respectively we obtain

$\frac{1}{T_i} = \left.\frac{\partial S_{\vec{k}}^{macro}}{\partial k_{i=1,\ldots,n-1}}\right|_{\vec{k}^0} \approx \frac{\partial S^{equil}}{\partial k^0_{i=1,\ldots,n-1}}$. This is an analog of the energy-balance equation – the balance

equation $\sum_{i=1}^{n-1} \frac{dk_i^0}{T_i} = dS^{equil}$. This equation could be obtained directly by the differentiating the logarithm of eq. (4).

For non-equilibrium states instead, the increment of entropy is $dS \geq d \ln N_{\vec{k}} = \sum_{i=1}^{n-1} dk_i \cdot \ln \frac{k_n}{k_i}$. It is easy to see that for the equilibrium increment of entropy this formula gives the previous equation $dS^{equil} = \sum_{i=1}^{n-1} dk_i^0 \cdot \ln \frac{k_n^0}{k_i^0} = \sum_{i=1}^{n-1} \frac{dk_i^0}{T_i}$.

Imagine now a system in the canonical ensemble during its evolution over non-equilibrium macrostates $\vec{k}$ on its way to the equilibrium. Each macrostate $\vec{k}$ could be thought as a system isolated with $\vec{k}$. Therefore the increase of the entropy in the system in accordance with eq. (7) must be higher than the increase of the entropy produced only by the change of macrostates (14): $dS_{\Delta N(\vec{k})}^{macro} \geq dS_{\vec{k}}^{macro} = d \ln N_{\vec{k}}$. In the vicinity of the maximum of $W^{equil}(\vec{k})$ we have eq. (15) and $dS \geq d \ln N_{\vec{k}} = \sum_{i=1}^{n-1} \frac{dk_i}{T_i}$. This gives $dA \equiv d\left(\sum_{i=1}^{n-1} \frac{k_i}{T_i} - S\right) = \sum_{i=1}^{n-1} \frac{dk_i}{T_i} - dS \leq 0$. So, we have confirmed that the Helmholtz free energy is a ruling potential.

## 5. Conclusion

An introduction of the stochasticity as an external model constraint introduces a noise into a system. For three model systems it is shown that this noise 'statistically' thermolizes in general a non-thermodynamic system. The equilibrium 'canonical' distribution of probabilities is dictated by the 'statistical' noise instead of the temperature of the external media. The formalism of the classical statistical mechanics can be further developed for the multifractal, non-Hamiltonian dynamical systems providing all classical features like the narrow probability maximum, the ruling (governing, free energy) potential, the balance equation, and the equation of state.